\documentclass[apl,preprint,superscriptaddress]{revtex4-1}

\usepackage[english]{babel}

\usepackage{graphicx}
\begin{document}

\title{
A scanning microcavity for in-situ control of single-molecule emission\\
}

\author{C. Toninelli}
\affiliation{Laboratory of Physical Chemistry and optETH, ETH
Zurich, CH-8093 Zurich, Switzerland}
\author{Y. Delley}
\affiliation{Laboratory of Physical Chemistry and optETH, ETH
Zurich, CH-8093 Zurich, Switzerland}
\author{T. St\"oferle}
\affiliation{IBM Research GmbH, Zurich Research Laboratory,
S\"aumerstrasse 4, 8803 R\"uschlikon, Switzerland}
\author{A. Renn}
\affiliation{Laboratory of Physical Chemistry and optETH, ETH
Zurich, CH-8093 Zurich, Switzerland}
\author{S. G\"{o}tzinger}
\affiliation{Laboratory of Physical Chemistry and optETH, ETH
Zurich, CH-8093 Zurich, Switzerland}
\author{V. Sandoghdar}
\affiliation{Laboratory of Physical Chemistry and optETH, ETH
Zurich, CH-8093 Zurich, Switzerland}
%

\date{\today}

\begin{abstract}
We report on the fabrication and characterization of a scannable
Fabry-Perot microcavity, consisting of a curved micromirror at the
end of an optical fiber and a planar distributed Bragg reflector.
Furthermore, we demonstrate the coupling of single organic molecules
embedded in a thin film to well-defined resonator modes. We discuss
the choice of cavity parameters that will allow sufficiently high
Purcell factors for enhancing the zero-phonon transition between the
vibrational ground levels of the electronic excited and ground
states.
\\

\end{abstract}

\maketitle


Single optical emitters such as atoms, ions, quantum dots, and color
centers have been heavily pursued for their potential in engineering
and control of the quantum state of matter. Among these, solid-state
emitters are particularly interesting because they are robust and
compatible with dense circuits. However, fabrication of high-quality
samples faces issues such as material processing when they are to be
integrated into more sophisticated structures. Recent
efforts~\cite{Kiraz:05,Gerhardt:09,Hwang:09} have shown that organic
dye molecules might offer a superior alternative if one could
suppress the incoherent decay of the excited state. In this work, we
present a versatile microcavity architecture that can enhance the
Fourier-limited transition between the v=0 vibrational levels of the
electronic excited and ground states (labeled $\gamma_{0-0}$ in Fig.
1(a)), thus reducing the influence of the broad Stokes-shifted
transitions (labeled $\gamma_{\rm Stk}$).

Over the past three decades, microresonators have been increasingly
employed for tailoring the interaction of light and matter
\cite{Vahala-book}. In the case of emitters embedded in solid-state
microcavities one has been usually confronted by two main problems.
First, it is not possible to combine arbitrary choices of emitters
and cavity material. Second, it is not easy to optimize the strength
of the coupling between the emitter and the cavity mode because of
the challenge in placing the former in the mode maximum with
subwavelength accuracy
\cite{Guthohrlein2001,Hennessy2007,Muller2009}. One way to address
these issues is to exploit evanescent coupling of an emitter that
has been placed at the extremity of a
nano-probe~\cite{Goetzinger:06,Koenderink:05b}. Here we explore
another flexible approach, where the emitters are embedded in a thin
film on a flat distributed Bragg reflector (DBR). By approaching a
microscopic mirror at the end of an optical
fiber~\cite{Trupke2005,Colombe2007}, we form a resonant microcavity
that can be laterally scanned to couple to individual emitters in a
controlled fashion~\cite{Muller2009}. This scheme is particularly
attractive for work with organic dye molecules, which have been so
far only studied in the near field of surfaces~\cite{Steiner:05,
Buchler:05} or in parallel-plate
cavities~\cite{Demartini1987,Steiner:05,Chizhik2009}.

\begin{figure}
\begin{center}
\includegraphics[width=0.47\textwidth]{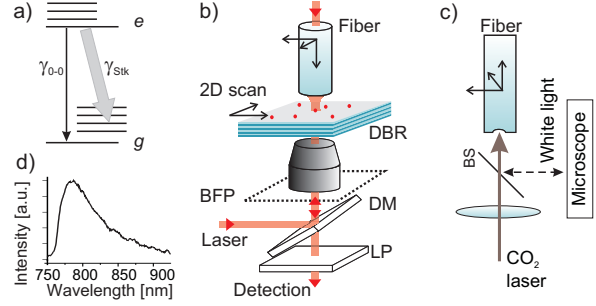}
\caption{(a) Schematics of the electronic states and vibrational
levels of an organic dye molecule. (b) Sketch of the cavity assembly
and experimental setup. Dichroic mirror (DM); long pass filter (LP);
the back focal plane of the objective (BFP). (c) Setup for laser
surface machining of fiber tips. The process is monitored with a
white-light microscope, coupled to the optical path through a beam
splitter (BS). (d) Fluorescence spectrum of an unperturbed single
DBT molecule.}\label{setup}\end{center}
\end{figure}

The schematics of the microcavity assembly is depicted in
Fig.\,\ref{setup}(b). One cavity mirror consists of a planar DBR
with alternating $\rm SiO_{2}$ and $\rm TaO_{5}$ bilayers of
$\lambda/4$ optical thickness. The lowest refractive index material
has been configured at the top interface to place the field antinode
at the emitter position. In our current experiments, the photonic
bandgap has been centered either on the emitter fluorescence or on
the excitation wavelength.

The curved microscopic mirror is formed at the end of a single-mode
optical fiber by laser machining, as shown in Fig.\,\ref{setup}(c).
By focussing a $\rm CO_2$ laser beam onto the cleaved fiber and
regulating its intensity in short pulses, we induce local
evaporation of silica and produce concave surfaces with variable
radii of curvature down to several tens of
microns~\cite{Colombe2007}. We investigated the laser processed
surfaces with atomic force microscopy (AFM) and found a rms
roughness below $1\,\rm nm$, limited by our instrument resolution.
We typically obtained a depression of $200\,\rm nm$ over a diameter
of $30\,\rm \mu m$ when the fiber was treated by $200$ ms bursts of
$8.5\, \rm \mu s$ pulses at a repetition rate of $5\,\rm kHz$,
corresponding to an average laser output power of about $200\,\rm
mW$. Longer pulses yield higher radii of curvature and deeper
concavities. Fiber tips were then coated with $100\, \rm nm$ of
gold. While this metallic coating reduces the finesse, it offers a
broadband response that was desirable in our current studies.
Extension to highly reflective dielectric coatings has been already
reported~\cite{Colombe2007,Muller2009}.

\begin{figure}
\begin{center}
\includegraphics[width=0.47\textwidth]{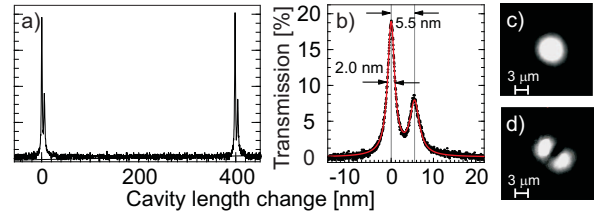}
\caption{(a,b) Laser transmission from the microcavity as the
micromirror is displaced in the axial direction. The red curve in
(b) is a double Lorentzian fit to the experimental data. (c) and (d)
are CCD images of the modes, corresponding to the two transverse
cavity modes in (a,b).}\label{FSR}\end{center}
\end{figure}

To characterize the resulting microcavity, we coupled light from a
single-mode diode laser at $780\,\rm nm$ to the cavity through its
optical fiber port. Light was then collected through a 60X
microscope objective and directed to a photodiode or a CCD camera.
Stable cavity modes could be found for several values of the cavity
length $L$. Figure\,\ref{FSR}(a) displays the cavity transmission
over more than one free spectral range (FSR) as $L$ was scanned.
These data yield typical cavity finesse of $\mathcal{F}=200$
according to $\mathcal{F}=FSR/\delta \nu= \lambda/2\delta L$, where
$\nu$ denotes the resonance frequency, $\delta \nu$ stands for the
cavity linewidth, and $\delta L$ signifies the corresponding length
change. In a separate measurement we determined the reflectivity of
the DBR (13 multilayers) to be $R_2=99.9\%$ at normal incidence. By
using this information, we deduced the reflectivity of the gold
micromirror to be $R_1=97\%$ from the expression
$\mathcal{F}=\frac{\pi \sqrt[4]{R_1R_2}}{1-\sqrt{R_1R_2}}$. We also
note that the measured cavity transmittance of $20\%$ is in
agreement with the predictions of simple plane-wave calculations,
confirming a good fiber-cavity mode matching.

The FSR of our short cavity amounts to about 100 nm, which is more
than the band gap of the DBR. Thus, to determine $L$, we varied the
cavity length under illumination at three different wavelengths of
$763\,\rm nm,\,775\,nm$ and $785\,\rm nm$. In this way, we obtained
an accurate calibration of the piezo displacement and determined
$L=m\lambda/2=2.75\,\rm \mu m$, corresponding to an effective
longitudinal cavity order of $m=7$. Accounting for the group delay
within the DBR of about $0.5\,\rm \mu m$, we estimate the physical
separation of the cavity mirrors to be $2.25\,\rm \mu m$.

Figure\,\ref{FSR}(b) shows the zoom of one longitudinal mode,
revealing two transversal resonances. Figures\,\ref{FSR}(c) and (d)
present CCD camera images of the cavity transmission at these
resonances, which we attribute to Hermite modes ($\rm TEM_{pn}$)
with indices $n=p=0$ and $n+p=1$, respectively. The difference
between two cavity lengths ($\Delta L$) can be used to determine the
radius of curvature ($r_1$) according to the relation $\Delta L =
\frac{\lambda}{2\pi}\Delta(n+p)\sqrt{\frac{L}{r_1}}$
~\cite{Yariv-book}. Assuming $L=7\lambda/2$, one obtains $r_1=1.4
\,\rm mm$ for this realization.

\begin{figure}
\begin{center}
\includegraphics[width=0.47\textwidth]{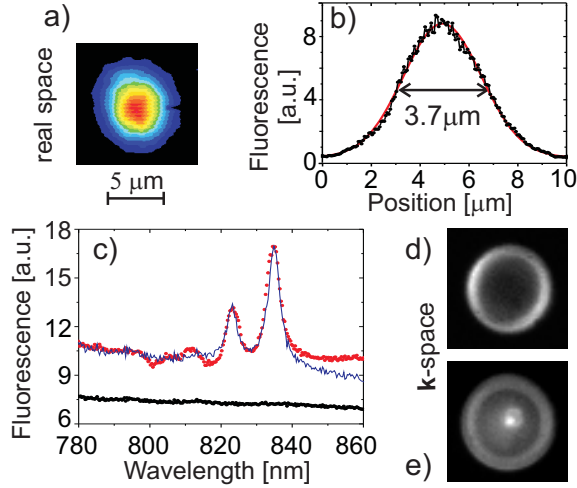}
\caption{(a) Fluorescence from a nanosphere as a function of its
lateral position in the cavity mode. (b) Horizontal cut through the
image in (a) (black dots) and Gaussian fit (red solid line). (c)
Emission of a single DBT recorded through the cavity (red dots) or
only through the bottom DBR (black dots). The blue trace is a fit
resulting from the product of a cavity transmission spectrum and
that of unperturbed DBT. (d, e) Back focal plane images of
fluorescence in the absence of the micromirror (d) and on resonance
with the cavity (e).} \label{real-k-space-fit}
\end{center}
\end{figure}

A Gaussian fit to the measurement in Fig.\,\ref{FSR}(c) yields a
full width at half-maximum (FWHM) of $3.8\,\rm \mu m$ at the planar
mirror surface. To map the cavity mode directly, we scanned a
100~nm-diameter fluorescent polymer bead laterally and recorded its
emission through the lower mirror. Here, we chose a DBR made of $12$
$\rm SiO_{2}$ and $\rm TiO_{2}$ bi-layers that yielded a band gap in
the range 524-684 nm. The light from a diode laser at $655\,\rm nm$
was coupled to the cavity resonantly via the fiber and was used to
excite the bead. At the same time, the fluorescence of the bead
(peaked at $755\,\rm nm$) could traverse the DBR and be collected
with an oil immersion objective (100X,\, N.A.=1.4). To ensure that
the cavity stayed tuned to its resonance, we stabilized $L$ via a
side-of-fringe locking scheme. Figure\,\ref{real-k-space-fit}(a)
presents an example of a two-dimensional image as the sample was
moved over a range of $100\,\rm \mu m^2$ at a step size of $250\,\rm
nm$. Since we kept the excitation intensity well below saturation,
this signal is proportional to the cavity field intensity. We find a
$\rm TM_{00}$ mode with $\rm FWHM=3.7\, \mu m$ (see
Fig.\,\ref{real-k-space-fit}(b)), consistent with our CCD
estimation. The knowledge of the FWHM of the mode at the DBR allows
us to determine its 1/e half-width for the field amplitude ($w_0$)
and thus compute the mode volume according to $V=\pi
w_0^{2}L/4\approx70\lambda^3$, taking into account the standing-wave
nature of the mode.

Having characterized the microcavity, we now examine its coupling to
a single Dibenzoterrylene (DBT) molecule. The ideal sample for our
microcavity concept is a thin dielectric film that contains organic
dye molecules with transition dipoles in the mirror plane. Recently,
we demonstrated that anthracene films (thickness 20-50 nm) doped
with DBT satisfy this requirement and provide a well-defined and
photostable system for single-molecule studies~\cite{Toninelli:10}.
We, thus, spin coated such a sample directly on a DBR made of 4
bilayers of $\rm SiO_{2}$ and $\rm TaO_{5}$ with band gap in the
range 685-880 nm. The excitation light at $720\,\rm nm$ experienced
only a weak cavity effect so that it could be focussed through the
DBR onto the molecule. The red trace in
Fig.\,\ref{real-k-space-fit}(c) plots a zoom into the fluorescence
spectrum of a single molecule recorded through the cavity.
Comparison with the emission spectrum of an unperturbed DBT molecule
in Fig.~\ref{setup}(d) reveals a clear modification due to the
coupling to two cavity transversal modes. To explore this effect
further, we also recorded the angular distribution of the single
molecule emission by imaging the back focal plane of the microscope
objective~\cite{Toninelli:10,Lieb2004} onto a sensitive CCD camera
(see Fig.~\ref{setup}(b)). Figure\,\ref{real-k-space-fit}(d)
displays the resulting pattern when the micromirror was absent. The
bright ring shows that in this case only the emission at large
angles, which falls outside the DBR band gap, can be transmitted.
The black symbols in Fig.~\ref{real-k-space-fit}(c) present part of
the fluorescence spectrum emitted at these angles.
Figure~\ref{real-k-space-fit}(e) shows that if $L$ fulfills a
resonant condition, the molecular fluorescence can build up in the
cavity and exit along its axis, resulting in a central peak in the
Fourier plane.

The data presented in Figs.~\ref{real-k-space-fit}(c)-(e) clearly
show that the molecular emission has been coupled to well-defined
modes of the microcavity. However, contrary to what has been
recently alluded in a related experiment~\cite{Chizhik2009}, such
observations cannot be attributed to a modification of the emission
spectrum arising from a Purcell effect. Instead, they are simply
caused by a filtering process, whereby the part of the molecular
emission that is not resonant with the cavity does not exit. To
confirm this, we multiplied the emission spectrum of DBT (see
Fig.~\ref{setup}(d)) by the transmission function of a cavity with
$\mathcal{F}=200$. To account for a small contribution from the
light exiting the DBR at large angles (see the ring in
Figs.~\ref{real-k-space-fit}(d)), we added a background according to
the experimental spectrum recorded through a DBR alone. As the blue
trace in Fig.~\ref{real-k-space-fit}(c) shows, this simple procedure
yields a very good agreement with the measured data of the red trace
without the need for invoking an intrinsic modification of the
molecular fluorescence.

Parameters $\mathcal{F}=200$, $L=7\lambda/2$, and $V=70 \lambda^3$
yield a maximal Purcell factor of about 1.5 for a narrow-band
emitter in a closed cavity according to the expression $
3Q\lambda^3/4\pi^2V $~\cite{Yokoyama-book}, where $Q$ is the quality
factor. Taking into account a solid angle of $0.3\times10^{-2}$
subtended by the cavity mode and its spectral overlap of $10^{-2}$
with the broad molecular emission, the modification factor drops to
about $5\times10^{-5}$. Given such a small change in the radiative
rate of the 0-0 transition, we cannot expect a considerable effect
on the redistribution of the emission among various vibrational
levels (see Figs.~\ref{setup}(a,d)). To enter this interesting
regime, we plan to coat the micromirror by a multilayer dielectric,
which has been reported to yield a cavity finesse of about
$3.7\times 10^4$~\cite{Colombe2007}. Combined with a modest radius
of curvature of $100~\rm \mu m$, such mirrors would allow Purcell
factors up to $1000$. Accounting for the reduction of this effect
due to the finite modal solid angle, such a cavity would enhance the
0-0 spontaneous emission by about 20 times and improve the fraction
of the emission in this channel from 30\%~\cite{Trebbia2009} to
about $85\%$ if one operates at liquid helium temperatures, where
0-0 linewidths under 1~GHz are
common~\cite{Kiraz:05,Gerhardt:09,Hwang:09}. Further reduction of
the radius of curvature of the fiber mirror would lower the mode
volume and allow branching ratios as large as 98\%. Such large
effects will also be accompanied by an improvement of the collection
efficiency through the resonator.

We thank R. Stutz for sputter deposition of DBRs. This work was
financed by ETH Zurich (QSIT, Grant Nr.~PP-01 07-02) and the Swiss
National Foundation.


%

\end{document}